\documentclass[aps,superscriptaddress,twocolumn,showpacs,preprintnumbers,amsmath,amssymb]
{revtex4}
\usepackage{ae,graphicx}

\newcommand{\be}{\begin{equation}}
\newcommand{\ee}{\end{equation}}
\newcommand{\ba}{\begin{eqnarray}}
\newcommand{\ea}{\end{eqnarray}}

\begin{document}

\title{Anisotropic Heisenberg model on hierarchical lattices with aperiodic 
interactions: a renormalization-group approach}

\author{N. S. Branco}\email{nsbranco@fisica.ufsc.br}
\affiliation{Departamento de Física,
Universidade Federal de Santa Catarina, 
88040-900, Florianópolis, SC, Brazil} 
\author{J. Ricardo de Sousa }
\affiliation{Departamento de Física,
UFAM, 3000-Japiim,
69077-000 Manaus, AM, Brazil}
\affiliation{Departamento de Física,
ICEx, Universidade Federal de Minas Gerais,
30123-970, Belo Horizonte-MG, Brazil}
\author{Angsula Ghosh}
\affiliation{Departamento de Física,
UFAM, 3000-Japiim,
69077-000 Manaus, AM, Brazil}

\date{\today}

\begin{abstract}

         Using a real-space renormalization-group approximation, we study the 
anisotropic quantum Heisenberg model on hierarchical lattices, with interactions following 
aperiodic sequences. Three different sequences are considered, with relevant and irrelevant
fluctuations, according to the Luck-Harris criterion. The phase
diagram is discussed as a function of the anisotropy parameter $\Delta$
(such that $\Delta=0$ and $\Delta=1$ correspond to the isotropic
Heisenberg and Ising models, respectively). We find three different types of
phase diagrams, with general characteristics: the isotropic Heisenberg plane is always an
invariant one (as expected by symmetry arguments) and the critical behavior of the 
anisotropic Heisenberg model is governed by fixed points on the Ising-model plane. Our results for
the isotropic Heisenberg model show that
the relevance or irrelevance of aperiodic models, when compared to their uniform counterpart, is
as predicted by the Harris-Luck criterion. A low-temperature renormalization-group procedure was applied
to the \textit{classical} isotropic Heisenberg model in two-dimensional hierarchical lattices:
the relevance criterion is obtained, again in accordance with the Harris-Luck criterion.

\noindent     

\end{abstract}
\pacs{}

\maketitle

\newpage

\section{Introduction}

		The investigation of systems displaying inhomogeneous or disordered interactions
is an active field of research \cite{cardy,seqs_aper}. From the experimental point of view, 
many of the materials found in
nature come with impurities; also, modern techniques are able to build
materials with controlled composition, such that two or more different atoms are
combined in a given order. Theoretically, one may be concerned with possible changes
on the critical behavior of systems, with the introduction of random disorder or inhomogeneuos
deterministic interactions, when compared to their homogeneous counterpart \cite{cardy}. 
For quenched random disorder, the Harris criterion \cite{harris} states that, if the pure-system's 
specific-heat exponent, $\alpha$, is positive (negative), the critical behavior of the disordered 
model is diffferent  from (the same as for) the pure model. 

	The discovery of quasi-crystals \cite{quasicristais} has motivated 
an intense research on the behavior of models with interactions following aperiodic sequences
\cite{seqs_aper}: numerical \cite{bercher} as well 
as analytical results \cite{vieira1,vieira2} have been obtained. Many works concentrated on
classical models, like Ising and Potts \cite{haddad1,aglae} ones, but some attention has
been drawn to quantum models in one dimension \cite{vieira1,vieira2,karevski,hida,luck2,arlego}. 
In Ref. \onlinecite{jagannathan}
the ground-state properties of a two-dimensional quantum model have been analyzed.
However, no work has focused on the finite-temperature critical behavior of quantum models
in dimensions two or above, to the best of our knowledge.

	A convenient model to address the role played by quantum effects 
and aperiodicity on critical phenomena is the anisotropic Heisenberg one, 
with interactions following aperiodic and deterministic sequences. On the other
hand, the way 
these sequences are constructed and the idea behind renormalization-group 
calculations make hierarchical lattices a natural choice for the study.
Therefore, in this work we treat the anisotropic
ferromagnetic Heisenberg model on four different hierarchical lattices,
with different Hausdorff fractal dimenions. Three different aperiodic sequences 
are treated, corresponding to bounded and unbounded fluctuations. We
will study here mainly the relevance of the introduction of
aperiodicity, in the renormalization-group sense. This aspect is generally
addressed by the so-called Harris-Luck criterion \cite{luck1}.
According to this criterion, the relevance of a given
aperiodic sequence is connected to the crossover exponent, $\phi$, given by
\cite{harris,luck1}:
\be
\phi = 1 - d_a \nu_0 \left( 1 - \omega \right), \label{eq:crossoverexponent}
\ee
where $d_a$ is the dimension the aperiodic sequence acts on, $\nu_0$ is the
correlation lenght's
critical exponent of the pure model and $\omega$ is the fluctuation exponent.  This exponent
is defined through
$g \sim N^{\omega}$, where $g$ is the fluctuation in the number of a given letter
of the sequence (below, we discuss this point further). For $\phi>0$, the critical behavior of the
aperiodic and uniform models are in different universality classes. For $\phi<0$, both aperiodic
and uniform models have the same set of critical exponents. For the marginal case,
$\phi=0$, critical exponents depend on the ratio between the two interaction constants
\cite{igloi,faria}; we will not duscuss this case further.
Our choices of hierarchical lattices
and aperiodic sequences allow for values of $\phi$ greater os smaller than zero, as well
as for different fractal dimensions of the lattice. The aperiodicity is chosen such that
$d_a=1$ in all cases we study.

	This work is organized as follows. In the next section we review some
basic concepts of aperiodic sequences, which will be important to our work.
In Section \ref{model} we define the model and outline the real-space 
renormalization-group approach we
use and in Section \ref{results} we present our results. In Section \ref{summary} 
we discuss and summarize the results.

\section{Aperiodic sequences} \label{sec:sequences}

	The aperiodic sequences used in this work are obtained
by the iteration of substitutuon rules working on an alphabet.
Each letter of the alphabet is replaced by a sequence of letters
and words are formed at each stage. We will be concerned with a 
two-letter alphabet, $A$ and $B$, and the usual convention is that
the initial word is $A$. More specifically, we will study the
following aperiodic sequences:

\begin{description}

\item[(i)] $A \rightarrow AB; \;\; B \rightarrow AA$,
i.e., from a given word of the sequence, the next word is obtained by
substituting $AB$ for every letter $A$ and
$AA$ for every letter $B$ in the previous word. The first stagess of this
sequence are $A \rightarrow AB \rightarrow ABAA \rightarrow ABAAABAB
\rightarrow \ldots$. This is the so-called period-doubling sequence.

\item[(ii)] $A \rightarrow ABB; \;\; B \rightarrow AAA$. The first
stages of this sequence are: $A \rightarrow ABB \rightarrow ABBAAAAAA \rightarrow
\ldots$;

\item[(iii)] $A \rightarrow AAB; \;\; B \rightarrow AAA$. In this case, the first
words of the sequence are: $A \rightarrow AAB \rightarrow AABAABAAA \rightarrow
\ldots$.

\end{description}

	The geometrical characteristics of these sequences are obtained from
the substitution matrix ${\cal M}$, which connects the number of letters
$A$ and $B$ after one application of the iteration rule, namely:
\be
        \left( \begin{array}{c} 
               N_{A}^{(n+1)} \\
               N_{B}^{(n+1)} 
               \end{array} \right) 
               = {\cal M}
        \left( \begin{array}{c}
               N_{A}^{(n)} \\
               N_{B}^{(n)} \end{array} \right)   \label{eq:matriz}
\ee

	Therefore, after $n$ iterations the total number of letters in the word,
$N^{(n)}$, is given by:
$ N^{(n)} \equiv N_{A}^{(n)} + N_{B}^{(n)} = {\cal M}^n N^{(0)}$,
where $N_{A}^{(n)}$ and $N_{B}^{(n)}$ are the number of letters $A$ and $B$ after $n$
iterations, respectively, and $N_{A}^{(0)}=1$ and $N_{B}^{(0)}=0$ for the initial word.

	The substitution matrices for the three aperiodic sequences defined
above are:

\textbf{(i)} $ {\cal M} = \left( \begin{array}{cc}
                         1 & 2 \\
                        1 & 0 \end{array} \right); $

\textbf{(ii)} $ {\cal M} = \left( \begin{array}{cc}
                         1 & 3 \\
                         2 & 0 \end{array} \right); $

\textbf{(iii)} $ {\cal M} = \left( \begin{array}{cc}
                         2 & 3 \\
                         1 & 0 \end{array} \right). $

	The total number of letters grows exponentially with the number
of iterations $n$:
\be
 N \sim \lambda_1^n, \; n\rightarrow\infty, \label{eq:Ncomn}  
\ee
where $N \equiv \lim_{n\rightarrow\infty} N^{(n)}$ and $\lambda_1$  is the greater
eigenvalue of ${\cal M}$. For the three sequences studied in this work, 
this exponential growth is valid for all
$n$ and $\lambda_1=2$ for sequence $(i)$ and $\lambda_1=3$ for sequences $(ii)$
and $(iii)$. One can define the fluctuation in a given letter, say $A$,
as $g^{(n)} = N_{A}^{(n)} - p_{A} N^{(n)}$, where
$p_{A}$ is the fraction of letters $A$ in the infinite word, i.e, after $n$
applications of the iteration rules, with $n \rightarrow \infty$. 
The fractions $p_{A}$ and $p_{B}$ are proportional to
the first and second entries, respectively, of the eigenvector corresponding 
to the greater eigenvalue. It is possible to show that:
\be 
g \sim \lambda_2^n, n\rightarrow\infty, \label{eq:gcomn}
\ee
where $g \equiv \lim_{n\rightarrow\infty} g^{(n)}$ and $\lambda_2$ is the smaller eigenvalue 
of ${\cal M}$. Therefore, using Eqs. (\ref{eq:Ncomn}) and (\ref{eq:gcomn}), one can show that:
\be
    g \sim N^{\omega}, \;\; \omega = \frac{\ln |\lambda_2|}{\ln \lambda_1}.  \label{eq:omega}
\ee

	The exponent $\omega$ is crucial for the crossover exponent, as outlined in the previous section.
Its value is $w=0$, $\ln(2)/\ln(3)$ and $0$ for
sequences $(i)$, $(ii)$ and $(iii)$, respectively, as can be easily calculated from
their substitution matrices. We will disccus the results for the crossover exponent
in Section \ref{results}.

\section{Model and Formalism} \label{model}

\begin{figure} 
\includegraphics{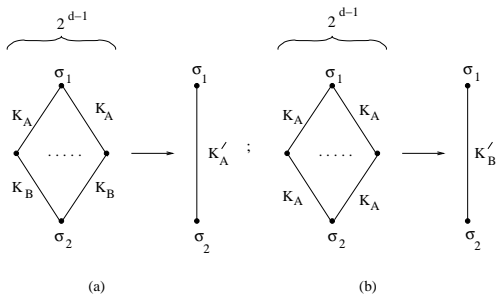}
\caption{\label{fig:redehierarqb2} Hierarchical lattice with $b=2$, suitable
for the study of the period-doubling sequence (sequence $(i)$; see text).
We show the renormalization for the coupling constant $K_A$ (part $(a)$)
and for the coupling constant $K_B$ (part $(b)$). Note that the construction
of the hierarchical lattice is made in the reverse order of the
renormalization-group procedure.}
\end{figure}

\begin{figure} 
\includegraphics{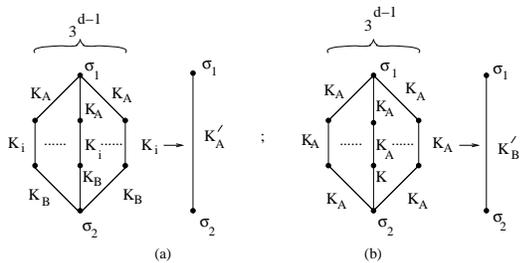}
\caption{\label{fig:redehierarqb3} Hierarchical lattice with $b=3$, suitable
for the study of sequences $(ii)$ and $(iii)$ (see text).
We show the renormalization for the coupling constant $K_A$ (part $(a)$),
where $K_i=K_B$ for sequence $(ii)$ and $K_i=K_A$ for sequence $(iii)$,
and for the coupling constant $K_B$ (part $(b)$).}
\end{figure}

	The reduced Hamiltonian of the anisotropic Heisenberg model is given by:
\be
  - \beta {\cal H} = \sum_{<i,j>} K_{ij} \left[ (1-\Delta) (\sigma_i^x \sigma_j^x 
+ \sigma_i^y \sigma_j^y) + \sigma_i^z \sigma_j^z \right],  \label{anisotropicHeisenberg}
\ee
where $\beta=1/k_B T$, $k_B$ being the Boltzmann constant and $T$ the temperature,
$\sigma_i^{\alpha}$ is the component $\alpha$ of a spin-$1/2$ Pauli
matrix on site $i$, $0 \leq \Delta \leq 1$ ($\Delta=0$ corresponds to the isotropic
Heisenberg model and $\Delta=1$ to the Ising model), the sum is over all 
first-neighbor bonds on a hierarchical lattice and the exchange constants
$K_{ij}= \beta J_{ij}$ follow an aperiodic sequence in a given direction of the lattice.
See Figs. \ref{fig:redehierarqb2} and \ref{fig:redehierarqb3} for examples of hierarchical
lattices with $b=2$ and $b=3$, respectively: these lattices are built of $b^{d-1}$ bonds
connected in parallel, each one consisting of $b$ bonds connected in series, where
$d$ is the fractal dimension of the lattice. In this work, we have treated lattices with $d=2$ and 
$d=3$.

	We use a
real-space renormalization-group approach; a partial trace is taken over internal spins
on suitable finite lattices and a renormalized Hamiltonian is obtained, namely:
\ba
  e^{- (\beta {\cal H})'}  & = & \exp \left\{ K_{12}' \left[ (1-\Delta') (\sigma_1^x \sigma_2^x 
+ \sigma_1^y \sigma_2^y) + \sigma_1^z \sigma_2^z \right] \right\} \nonumber \\
 & \equiv & \mbox{Tr}_{\{\sigma\}} e^{- \beta {\cal H}} ,   \label{eq:rg}
\ea
where $K_{12}'=K^{'}_A$ in Figs. \ref{fig:redehierarqb2}$(a)$ and
\ref{fig:redehierarqb3}$(a)$ and $K_{12}'=K^{'}_B$ in Figs.
\ref{fig:redehierarqb2}$(b)$ and \ref{fig:redehierarqb3}$(b)$,
$\mbox{Tr}_{\{\sigma\}}$ is a partial trace, taken over all spins in
Figs. \ref{fig:redehierarqb2} and  \ref{fig:redehierarqb3}, except
$\sigma_1$ and $\sigma_2$, and $\beta {\cal H}$ is the reduced Hamiltonian
of the cell on the left-hand sides of Fig. \ref{fig:redehierarqb2} $(a)$
and \ref{fig:redehierarqb2} $(b)$ and Fig. \ref{fig:redehierarqb3} $(a)$ and 
\ref{fig:redehierarqb3} $(b)$.
The method to calculate the partial trace in Eq. (\ref{eq:rg}) for quantum systems was introduced in
Ref. \onlinecite{mariz} and an important simplification was proposed 
in Ref. \onlinecite{sousa1}, where the whole process is explained in detail.
This method has been successfully applied in the study of ferromagnetic, antiferromagnetic
and spin-glass quantum models.  The formalism is specially suitable to obtain
multidimensional
phase diagrams and qualitative results, indicating universality
classes and possible crossover phenomena. It is worth mentioning
that, although $\Delta$ is initially uniform and the aperiodicity acts 
only on the interaction parameter $K$, after the first iteration of the 
renormalization-group the anisotropy is no longer the same for every bond.
This fact has to be taken into account when deriving the recursion relations.
We refer the reader to Refs. \onlinecite{mariz} and \onlinecite{sousa1} for
details.

\section{Results and Discussion} \label{results}

	Some features are common to all three sequences: the isotropic Heisenberg model
($\Delta=0$) is an invariant sub-space, in the renormalization-group sense. The same
applies for the Ising model ($\Delta=1$). The reason is that the introduction of exchange
constants with different values does not change the symmetry of these two models. Therefore, the flow
will not leave the corresponding subspaces. Moreover, the critical behavior for
$0 < \Delta < 1$ will be determined by the stability of the Ising-model non-trivial fixed points, i.e,
the flow for any initial value of $\Delta$ in that range is towards the $\Delta=1$ subspace.
The Ising model with aperiodic interactions has already been treated and our 
results for $\Delta=1$ agree with those in Ref. \onlinecite{haddad1}. Particularly, the stability of the
uniform fixed points, with respect to the introduction of aperiodicity, is in accordance
with the Harri-Luck criterion. Therefore, from now on we will
restrict ourselves to the isotropic Heisenberg model subspace.

\begin{figure}[ht] 
\includegraphics{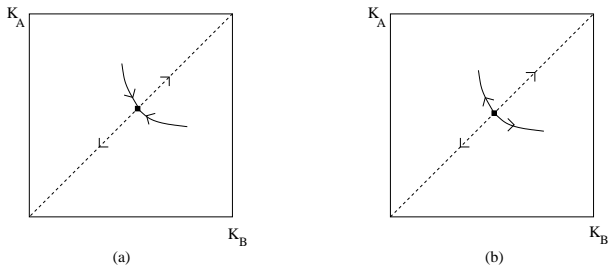}
\caption{\label{fig:fluxo} Qualitative phase diagram for hierarchical lattices with
$d=3$. $(a)$ Renormalization-group flux for irrelevant aperiodic 
sequences; $(b)$ Renormalization-group flux for relevant aperiodic 
sequences. The straight lines at $45$ degrees represent the uniform model, $K_A=K_B$,
and the fixed point is relevant along this direction.}
\end{figure}

	In the renormalization-group framework, the stability of fixed points is given by the eigenvalues of the
matrix of the linearized renormalization-group equations (LRGE) \cite{cardy}. Since the relevant fixed points
(filled squares in Fig. \ref{fig:fluxo}) are the non-trivial ones, one of these eigenvalues is always greater 
than one and
corresponds to the flux along the uniform sub-space (traced lines in Fig. \ref{fig:fluxo}). In this figure
we show the qualitative picture we obtain for irrelevant (part $(a)$) and relevant (part $(b)$) aperiodic 
sequences for $d=3$.
The uniform model corresponds to the straight line at $45^o$, where $K_A=K_B$; the fixed point is
always unstable along this line. The relevance of the aperiodicity is given by the stability
along the other direction (continuous lines leaving the fixed points in Fig. \ref{fig:fluxo}): in
part $(a)$ of the figure, the aperiodicity does not change the critical behavior, compared to the
uniform model, while in part $(b)$ a new universality class emerges when aperiodicity is introduced.
Note that the phase diagrams for lattices with $d=2$ are qualitative different from the ones
in Fig. \ref{fig:fluxo}: the non-trivial fixed points are at zero temperature ($K=\infty$) and,
therefore, the ''aperiodic'' direction is not physically accessible in these cases.
Nevertheless, the relevance of the aperiodicity is correctly described by the renormalization-group
formalism, as we will see below.

	A technical point is worth mentioning here.
For all sequences we treat in this work the structure of the matrix of the LRGE,
evaluated at the uniform fixed point, $K^{*} \equiv K_A^{*} = K_B^{*}$, is:
\begin{equation}
\left(
\begin{array}{cl}
 \left. \frac{\partial K^{'}_A}{\partial K_A}\right|_{K^{*}} & \left. \frac{\partial K^{'}_A}{\partial K_B}\right|_{K^{*}} \nonumber \\
 \left. \frac{\partial K^{'}_B}{\partial K_A}\right|_{K^{*}} & \left. \frac{\partial K^{'}_B}{\partial K_B}\right|_{K^{*}}=0
\end{array}
\right) ,
\end{equation}
Since:
\begin{equation}
 \left. \partial K^{'}_B / \partial K_A \right|_{K^{*}} = 
\left. \partial K^{'}_A/ \partial K_A \right|_{K^{*}} + \left. \partial K^{'}_A/ \partial K_B \right|_{K^{*}},
\end{equation}
the eigenvalues of the above matrix are:
\begin{equation}
\Lambda_1=\left. \frac{\partial K^{'}_B}{\partial K_A}\right|_{K^{*}}; 
\Lambda_2 = -\left. \frac{\partial K^{'}_A}{\partial K_B}\right|_{K^{*}}
\end{equation}
The former corresponds to the uniform model,
as discussed previously, and, therefore, is always greater than 1. 
The absolute value of the latter eigenvalue determines the
relevance of the aperiodicity.

	For classical models, like the Ising and Potts ones \cite{haddad1}, the matrix
of the LRGE is proportional to the transpose of the substitution matrix 
(see Eq.  (\ref{eq:matriz}) above), ${\cal M}^{\mbox{T}}$. This property still holds
true for quantum models \textit{and} for aperiodic sequences such that the hierarchical
lattices which renormalizes into $K^{'}_A$ is symmetric with respect to the exchange of the iterations 
$K_A$ and $K_B$, as in part $(a)$ of Fig. \ref{fig:redehierarqb2}.
This brings an important simplification for the calculation
of the former matrix, since the proportionality factor can be obtained from the 
recursion relation for the \textit{uniform} model, i.e, from the renormalization of $K_B$. 
Then, one can obtain the other two elements, $\left. \partial K^{'}_A/ \partial K_A \right|_{K^{*}}$
and $\left. \partial K^{'}_A/ \partial K_B \right|_{K^{*}}$,
from the substitution matrix, with no need to work out the recursion 
relation for the aperiodic model (this relation involves bonds with different values,
which may make it cumbersome to be calculated). Note that, while the proportinality
between the matrix of the LRGE and ${\cal M}^{\mbox{T}}$
is always true for classical models,
it fails for quantum models when the lattice is not symmetric with
respect to the exchange of the iterations  $K_A$ and $K_B$ (like the one in part $(a)$ of 
Fig.  \ref{fig:redehierarqb3}, for example).

	For the hierarchical lattices with $b=3$, we have to resort to the Migdal-Kadanoff
approximation, since the number of sites involved in the renormalization is rather large.
This approximation is equivalent to treating
the cell ``by pieces'', i.e., to renormalizating first the $b$ bond in series and then
combining the $b^{d-1}$ renormalized bonds in parallel. An alternative (and more precise, since
commutation aspects are taken into account at the cell level) procedure
is to renormalize the whole cell. This point is extensively discussed in Ref. \onlinecite{mariz},
where it is shown that treating the cell ``by pieces'' gives good qualitative results for
ferromagnetic quantum models, when compared to renormalizing the whole cell. The comparison 
between these two procedures has also been made for antiferromagnetic
quantum models \cite{branco2} and again the agreement is quite good (note, however, that for
models such that frustration effects are present, the two approaches give different
qualitative results \cite{sousa2})). In this work, we have applied the two procedures outlined above
for $b=2$; the qualitative (and sometimes even the quantitative) agreement is excellent and put 
the Migdal-Kadanoff approximation made for $b=3$ on a firmer basis.

	The hierarchical lettices we treat here are those with $b=2$, $d=2$ and $3$,
(see Fig.~\ref{fig:redehierarqb2}) and $b=3$, $d=2$ and $3$ (see Fig.~\ref{fig:redehierarqb3}).
The corresponding results are:

\begin{itemize} 

\item[(a)] $b=2, d=2$: in this case, the results are the same treating the cell as
a whole or within the Migdal-Kadanoff approximation. The critical temperature, $T_c$, and the
correlation-length's critical exponent, $\nu$, are known exactly for the two-dimensional
uniform model, namely $T_c=0$ \cite{mermin} and $\nu=\infty$.
We obtain these exact results with our procedure. Therefore, according to the Harris-Luck criterion, 
Eq.~(\ref{eq:crossoverexponent}), the crossover exponent is negative for any sequence.
In fact, the eigenvalues for this model are:
$\Lambda_{1}=1$ and $\Lambda_{2}=-1/2$. The former corresponds to the
pure-model's critical behavior; the second determines the irrelevance of the aperiodicity,
as predicted by the Harris-Luck criterion. The negative sign of the second largest
eigenvalue is a signature of aperiodic systems 
\cite{haddad2};

\item[(b)] $b=2, d=3$: in this case, we have $I)$ renormalized the cell as a whole
and $II)$ used the Migdal-Kadanoff approximation, as in the last item. In both
procedures we obtain a finite critical temperature ($T_c=2.70$ for the former
and $T_c=2.91$ for the latter), as expected. The approximated value for the critical 
exponent $\nu$ is: $I)$ $1.511$, and $II)$ $1.398$. Since $\omega=0$ for the period-doubling
sequence, the crossover exponent is negative in both approximations, namely: $I)$ $\phi=-0.511$,
and $II)$ $\phi=-0.398$. Therefore, we expect the aperiodicity to be irrelevant.
The eigenvalues of the matrices of the LRGE are:
$I)$ $\Lambda_1 = 1.582$; $\Lambda_2 = -0.791$; and $II)$ $\Lambda_1 = 1.642$; $\Lambda_2 = -0.821$.
As expected, in both cases the aperiodicity is irrelevant and the smaller eigenvalue is
nagative, as commented above;

\item[(c)] $b=3, d=2$: again, $T_c=0$ and $\nu=\infty$ for the uniform model. Our procedure
(see Fig. \ref{fig:redehierarqb3}) obtains the correct critical temperature but the value
obtained for $\nu$ is an excellent approximation but not the exact one, namely
$\nu=8.494$. As discussed above, we have to resort to the Migdal-Kadanoff procedure, in this case,
but, in view of the comparison made in $(a)$ and $(b)$, we believe that the physical
behavior is correctly described by this approximation.
For this lattice, we have studied sequences $(ii)$ and $(iii)$ (defined in Section
\ref{sec:sequences}). For sequence $(ii)$, the wandering exponent is $\omega=\ln(2)/\ln(3)$ and
the crossover exponent is $\phi=-2.135$, i.e., the aperiodicity is irrelevant, according to
the Harris-Luck criterion; the situation is analogous for sequence $(iii)$, where 
$\omega=0$ and then $\phi=-7.494$: the aperiodicity defined by this sequence is also
irrelevant. For the eigenvalues of the substitutional matrices we obtain: 
$(ii)$: $\Lambda_1 = 0.879$; $\Lambda_2 = -0.654$;
$(iii)$: $\Lambda_1 = 0.879$; $\Lambda_2 = -0.224$. Therefore, the aperiodicity is irrelevant
for both cases, as predicted by the Harris-Luck criterion. Note that, since the relevant 
fixed point is at $K^{*}_{A}=K^{*}_{B}=\infty$, the value $\Lambda_1 = 0.879$ means
that this point unstable, in the renormalization-group sense;

\item[(d)] $b=3, d=3$: we have used the Migdal-Kadanoff approximation and obtained $T_c=1.92$ and
$\nu=1.551$ for the uniform case. Recalling the values for the wandering exponents
for sequences $(ii)$ and $(iii)$ (see $(c)$ above), the crossover exponents are given by
$\phi=0.427$ and $\phi=-0.551$, respectively. The eigenvalues of the matrix of the LRGE
are: $(ii)$: $\Lambda_1 = 2.030$; $\Lambda_2 = -1.363$;
$(iii)$: $\Lambda_1 = 2.030$; $\Lambda_2 = -0.667$.  Therefore,
sequence $(ii)$ is relevant and sequence $(iii)$ is irrelevant, again in
accordance with the Harris-Luck criterion. For the relevant case, a fixed
\textit{cycle} of period two emerges, as already found in Ref. \onlinecite{haddad2}.
The location of this stable two-cycle is $(K_{A}^{*}=0.417,K_{B}^{*}=3.90)$;
$(K_{A}^{*}=1.85,K_{B}^{*}=0.325)$. The eigenvalues of the LRGE associated with this
double iteration are $\Lambda_1 = 3.89$; $\Lambda_2 = 0.250$. The specific heat
critical exponent associated with the aperiodic fixed cycle, $\alpha_{a}$, is calculated 
from the relation:
\[ \alpha_{a} = 2 - d \frac{\ln b^2}{\ln \Lambda_1}, \]
where $d$ is the fractal dimension of the hierarchical lattice and $b$ is the scaling parameter
associated with onde iteration.
In this case, we obtain $\alpha_{a} = -2.855$, which is smaller than its counterpart for
the uniform model $\alpha_{u} = -2.653$ (this value can be obtained from the value
of $\nu$, quoted above, and the relation $\alpha_{u} = 2 - d \nu$).

\end{itemize}

	We have also studied the \textit{classical isotropic} Heisenberg model at low temperatures.
Only in this limit this model is closed upon application of the renormalization-group
transformation. Our procedure is appropriate to the study of two-dimensional systems,
since only in these cases the non-trivial fixed point is at zero temperature. 
The renormalized parameters are \cite{stinchcombe}:
\begin{equation}
 \frac{1}{K'_{A}} = \frac{1}{b} \left( \frac{n_1}{K_{A}} + \frac{n_2}{K_{B}} \right);
 \frac{1}{K'_{B}} = \frac{1}{b} \left( \frac{n_3}{K_{A}} + \frac{n_4}{K_{B}} \right) \label{eq:renormclassica}.
\end{equation}
These equations assume an aperiodic sequence built by the substitution rules:
\begin{equation}
 A \rightarrow \overbrace{AA \cdots A}^{n_1}\; \overbrace{B \cdots B}^{n_2};
\;\;\; B \rightarrow \overbrace{AA \cdots A}^{n_3}\; \overbrace{B \cdots B}^{n_4} \label{eq:seqgeral}.
\end{equation}
Note that, since the model is classical, the order of the interactions in Eqs. (\ref{eq:seqgeral})
is not relevant and the proportinality between the matrix of the LRGE and ${\cal M}^T$ holds in
this case. It is easy to show that, for $K^{*}_{A}=K^{*}_{B}=\infty$, the matrix of the
LRGE is given by:
\begin{equation}
      \left( \begin{array}{cc}
                b n_1 / (n_1+n_2)^2 & b n_2 / (n_1+n_2)^2 \\
                b n_3 / (n_3+n_4)^2 & b n_4 / (n_3+n_4)^2 \end{array} \right).
\end{equation}
Since $n_1+n_2=n_3+n_4=b$ and $n_4=0$ in the sequences studied here, the eigenvalues are
$\lambda_1=n_3/b=1$ and $\lambda_2=-n_2/b$, with $|\lambda_2|<1$. So, the
eigenvalues of the substutitional matrix are $\Lambda_1=b$ and $\Lambda_2=-n_2$,
and the wandering exponnet is given by:
\begin{equation}
            \omega= \frac{\ln n_2}{\ln b} < 1.
\end{equation}
Note that the value of $\lambda_1$ implies that $\nu=\infty$ and, since $\omega<1$,
any aperiodic sequence with $n_1+n_2=n_3+n_4=b$ will be irrelevant for the classical
isotropic Heisenberg model in two dimensions.  This result is supported by the formalims we
apply in this work, since $|\lambda_2|<1$.

\section{Summary} \label{summary}

	Witihin a real-space renormalization-group framework, we have studied the quantum
anisotropic Heisenbeg model with interactions following three different aperiodic sequences,
on  four different hierarchical lattices.
We obtain the exact result $T_c=0$ for two-dimensional lattices, while our evaluation of $T_c$
is always finite when $d=3$. In accordance with symmetry arguments, the isotropic Heisenberg-model 
subsapce in an invariant one and the flow of the anisotropic models is always towards the 
Ising subspace. Our procedure allows for the calculation of the stability of the
homogeneous fixed points, which agrees with the Harris-Luck criterion in all studied cases.
For the relevant sequence, we established the presence of a new stable fixed cycle of period two 
and calculated its sepcific heat critical exponent.
We also applied a low-temperature renormalization-group calculation to the isotropic classical
Heisenberg model in two-dimensional lattices: the results we obtain are exact on the
respective hierarchical lattices and on this range of temperatures. We obtain that,
on two-dimensional hierarchical lattices, all aperiodic sequences are irrelevant,
in agreement with the Harris-Luck criterion.

\begin{acknowledgments}
The authors would like to thank FAPESC, FAPEAM, CNPq, and CAPES for partial financial support.
\end{acknowledgments}

\bibliography{references}

\end{document}